\documentclass[twocolumn,showpacs,preprintnumbers,amsmath,amssymb,prc]{revtex4}
\usepackage{graphicx}   
\graphicspath{{./}{./figs/}}
\begin{document}
%
%
\title{Beam Test of the PIN-diode Readout Units with Electron Energies from 5 to 40 GeV at CERN SPS}
\author{Chengbo Li $^{1,2,3}$}
\thanks{Supported by National Natural Science Foundation of China (00121140488)) }
\email{lichengbo2008@163.com}

\author{Xiaomei Li $^{2}$}
\author{Qiuying Meng $^{2}$}
\author{Jing Zhou $^{2}$}
\author{Shuhua Zhou $^{2}$}
\author{Jian Yuan $^{2}$}

\affiliation{Key Laboratory of Beam Technology of Ministry of Education, Beijing Radiation Center, Beijing 100875, China}
\affiliation{Department of Nuclear Physics, China Institute of Atomic Energy, Beijing 102413, China}
\affiliation{College of Nuclear Science and Technology, Beijing Normal University, Beijing 100875, China}

\date{\today}
\begin{abstract}
The large-area violet-light-sensitive silicon photodiode PIN is one of the candidates of the lead tungstate crystal detector readout component of the photon spectrometer in the large heavy ion collision experiment. The PIN diode was assembled with the lead tungstate crystal and the low-noise preamplifier into a complete detector unit. The beam test was carried out on the SPS accelerator at CERN. The energy resolution was measured with the electron beam energy ranging from 5 to 40 GeV. The summation correction method was discussed, and an excellent linearity of the nominal beam energy versus the peak position of the detector was obtained, which showed the punch-through effect can be ignored.

\end{abstract}
%
\maketitle
\section{Introduction}

ALICE (A Large Ion Collider Experiment) \cite{alice} is an experiment at the LHC (Large Hadron Collider) optimized for the study of heavy-ion collisions, at a centre-of-mass energy about 5.5 TeV. 
The prime aim of the experiment is to study in detail the behaviour of matter at high densities and temperatures, in view of probing deconfinement and chiral-symmetry restoration.
The major technical challenge to the experiment is imposed by the large number of particles created in the collisions of lead ions. There is a considerable spread in the currently available predictions for
the multiplicity of charged particles produced in a central Pb-Pb collision. 
The detector consists essentially of two main components: the central part, composed of detectors mainly devoted to the study of hadronic signals and dielectrons in the pseudorapidity range $-1< \eta< 1$, and the forward muon spectrometer, devoted to the study of quarkonia behaviour in dense matter. 

The central part of the ALICE detector also includes an electromagnetic calorimeter, PHOS (PHOton Spectrometer)\cite{phos}, devoted to the study of photon signals.
It is positioned on the bottom of the ALICE set-up, and covers approximately a quarter of a unit in pseudorapidity, $-0.12 \leq \eta \leq 0.12$, and 100° in azimuthal angle. 
The PHOS electromagnetic calorimeter is optimized for measuring photons (of 0.5-10 GeV/$c$), $\pi^0$'s (of 1-10 GeV/$c$) and $\eta$ mesons (of 2-10 GeV/$c$). 
Measurement of higher momenta particles will also be feasible, although not with optimal energy resolution.
PHOS is consists of 17280 detecting channels of lead-tungstate crystals, PWO ($\rm PbWO_4 $), of $ \rm 2.2 \times 2.2 \times 18$ $\rm cm^3 $ dimensions, coupled to large-area PIN-diodes with low-noise preamplifiers.

The PIN diode and APD (avalanche photodiode) are designed as two candidates of the readout unit of PWO crystal in PHOS. The PIN diode has better linearity with large sensitive area, while the APD has higher transmission gain with small area.

PIN diode is a kind of large-area, thick depleted depth silicon photodiode, which can convert photons into electrons. Different PIN diode has different size and character. So, it has a wide range of applications\cite{pin89,pin93,pin13,pin15,pin18}, such as photon detecting, X-ray detecting, γ-ray detecting, microwave, antenna, mobile, radiotherapy, and scientific research et al.

The Chinese PIN diode was developed by CIAE (China Institute of Atomic Energy) and Peking university \cite{nt2006}. It has a large sensitive area of 16 x 17 $\rm mm^2$,  the junction capacitance is 110-120 pf when it is fully depleted, the leak current at room temperature is less than 5 nA, and the quantum efficiency in the violet light region (400-500 nm) is about 82\%. The noise of PIN+preamplifier is less than 527 ENC (equivalent noise charge) at -25℃.

After the offline basic tests of the main performance of Chinese PINs, they were taken to CERN and assembled into a complete PHOS detector unit (PWO + PIN + Preamplifier) to perform the beam test.

\section{Beam test}

\begin{figure}
\begin{center}
\includegraphics[width = 0.35\textwidth]{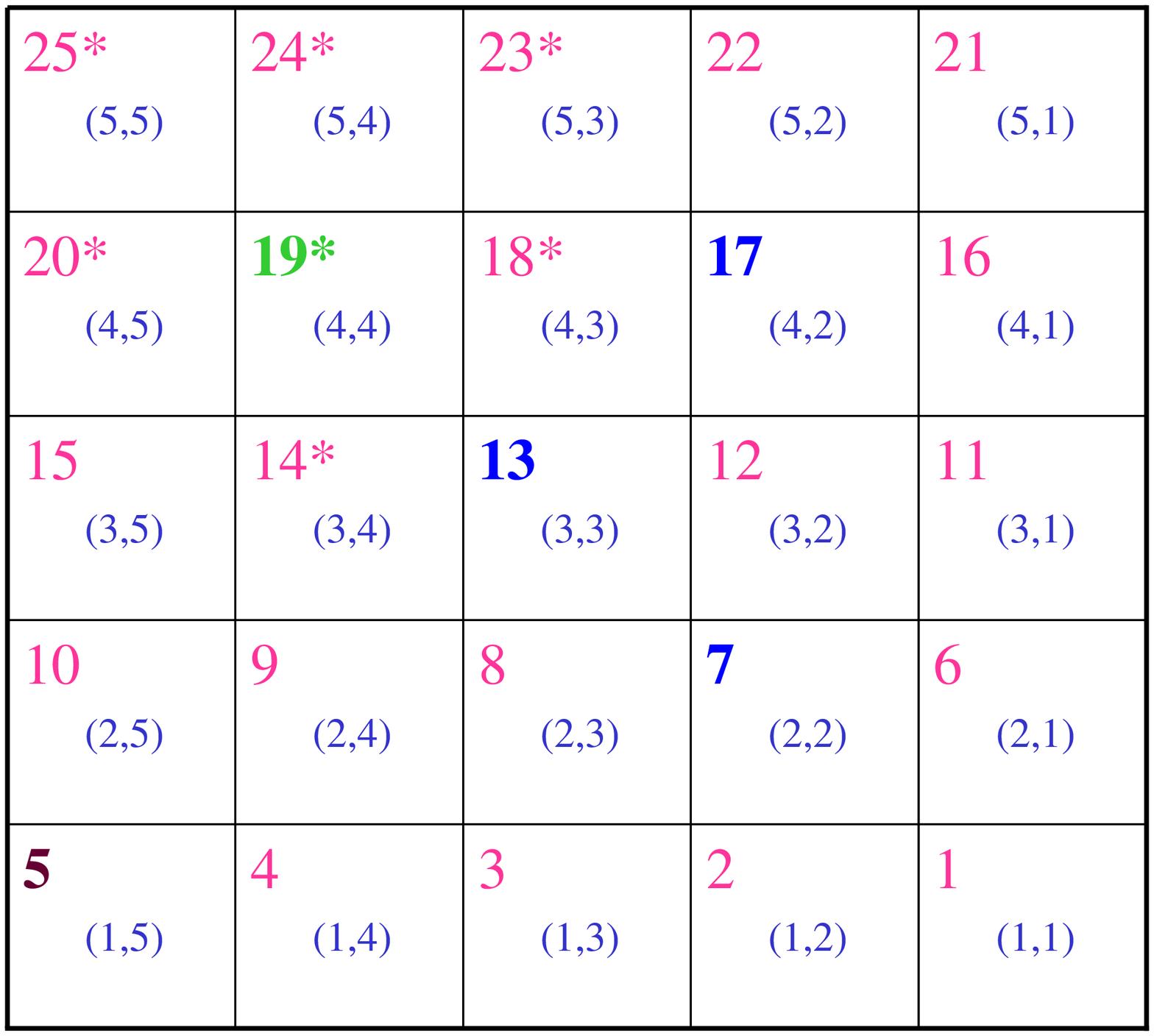}\\
\caption{(Color online) The 5x5 detector matrix. As a comparison, the 3x3 small matrix in the upper left corner are Norwegian PINs marked with *.}
\label{fig1} 
\end{center}
\end{figure}

The PIN diode connecting with the preamplifier is bonded to the bottom of the PWO using silica gel, so that the contact is good enough without bubbles, thus forming a complete detector unit. 25 detector units are arranged into a 5x5 detector matrix. As a comparison, the 3x3 small matrix in the upper left corner are Norwegian PIN samples marked with *, as shown in Figure \ref{fig1}.
The detector matrix is placed in the low temperature environment of -25℃, and installed on the movable platform of X5 beam line of SPS accelerator. Each detector unit can be located to the beam spot by remote adjusting along X and Y directions, so the experimental measurement can be carried out.

The electron beam provided by the SPS accelerator hits the PWO crystal of the detector, and fluorescences are produced by electromagnetic shower in PWO, then they are converted into photoelectrons by PIN. The electrical signals output by the preamplifier are sent to the forming amplifier, and then sent to the ADC. At the same time, a trigger signal given by the logic combination of the trigger detector signals  is sent to the ADC to open the ADC gate. The information of the ADC and the trigger register signal in another CAMAC crate are both recorded by a multi-parameter acquisition system.
The DAQ system 'CASCADE' is consists of CAMAC, VME, SCI high-speed transmission line and a computer with PCI card. The DAQ PC is working with PAW interface in Linux system.
And the temperature of the detector matrix is continuously monitored on a separate control PC.
All information is recorded on an event-by-event basis and written to tapes for offline analysis.

The main purpose of the experiment is to measure the energy resolution of the detector unit in a wide range of energy using the SPS electron beam. 
The experiment is arranged as follows:

(1) The 5x5 detector units are scaned respectively by a 20 GeV electron beam for energy calibration.

(2) The central detector of each 3x3 small matrix is scanned by electron beam with different energy from 5 to 40 GeV. 
Among them, the central detector of No.7 is measured in detail with more energy points, and others with less points.

\section{Data analysis and results}

In an ideal case, each detector of the photon spectrometer is an independent detector unit: the incident electrons produce showers in the crystal ($e^- \rightarrow \gamma \rightarrow  e^- e^+ \rightarrow   \gamma  \gamma ...  \rightarrow ...$ ). The electrons and photons will be finally converted to fluorescences in the crystal, and then the fluorescences are converted to electrical signals in PIN. 
In fact, due to the limit of crystal geometry size, some electrons or γ-rays of the shower may pass through the crystal when a high-energy electron hits on the detector. 
If the shower particles pass through the crystal longitudinally, they will enter the PIN diode. Because the PIN diode is not only sensitive to photons, but also very sensitive to charged particles, it will cause the distortion of the output signal, this is so-called the punch-through effect. 
If the shower particles pass through the side of the central crystal and enter other crystals around, the amplitude of the real signal will be reduced, which can be corrected by a summation method:
\begin{equation}\label{eq1}
 A_{c,sum}=A_c + \sum_i \eta_{i,c} \cdot A_i
\end{equation}
\begin{equation}\label{eq11}
 \eta_{i,c}= \frac{Peak_c}{Peak_i}
\end{equation}
Where $A_c$ is the signal amplitude generated by an event in the central detector, $A_i$ is the signal amplitude caused by the same event in the detector around it, and $A_{c,sum}$ is the corrected central detector signal. Where $\eta_{i,c}$ is the sum-factor, which is the ratio of peak position of the central detector with other detector around caused by the scaning electron beam with the same energy.

The sum-number should be determined according to the actual situation of the experimental setup. If the sum-number is too small, some useful signal will be lost, while if the sum-number is too large and too much noise will be introduced, the energy resolution will also be reduced. 
The usual sum-number is as follow: 

(1) Sum(1=center): only the single central detector is used.

(2) Sum(5=1+4): the central detector and the upper, lower, left and right detectors.

(3) Sum(9=3x3): the 3x3 matrix around the central detector.

(4) Sum(25=5x5): the 5x5 matrix around the central detector.

\begin{figure}
\begin{center}
\includegraphics[width = 0.4\textwidth]{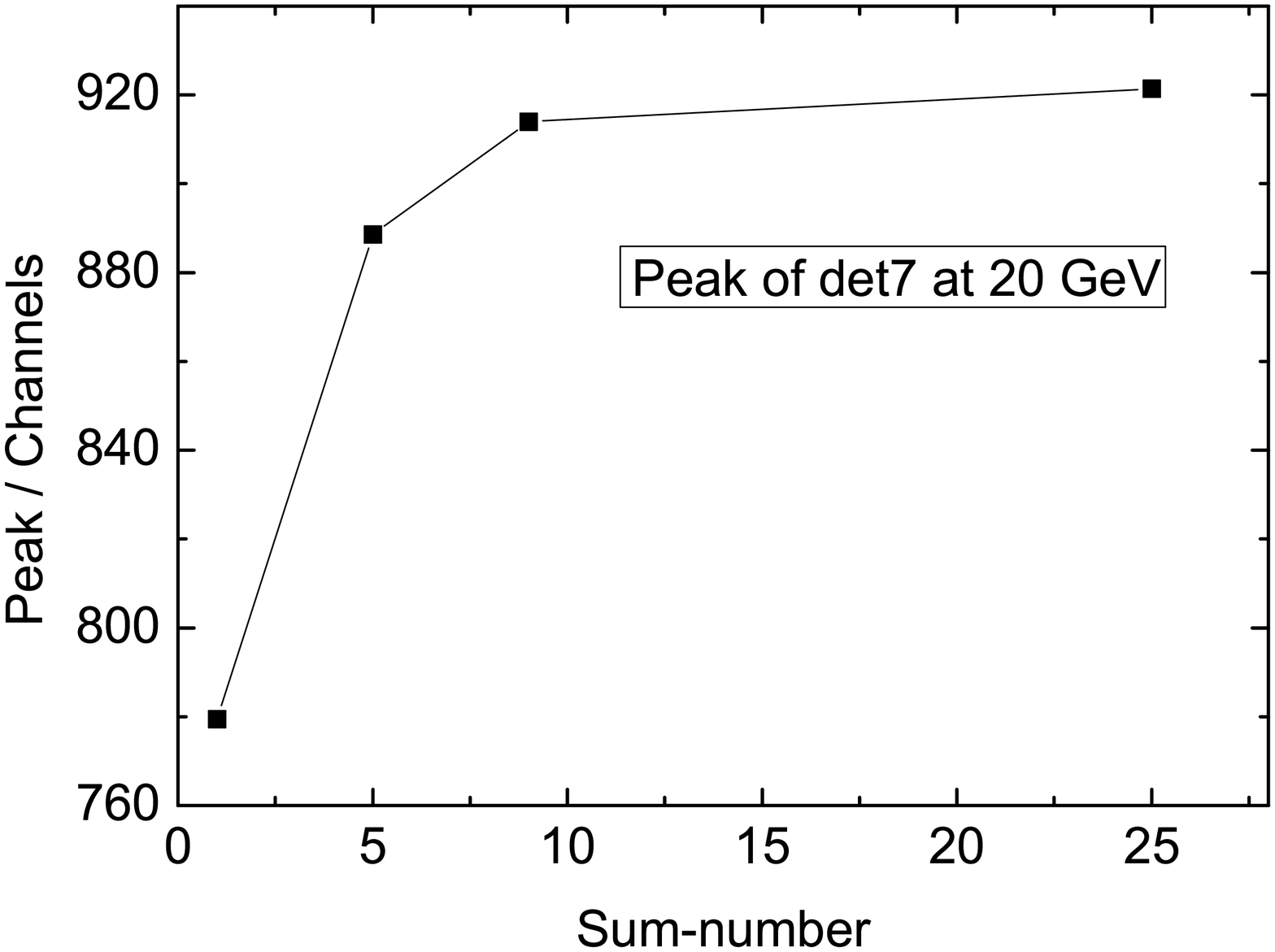}\\
\includegraphics[width = 0.4\textwidth]{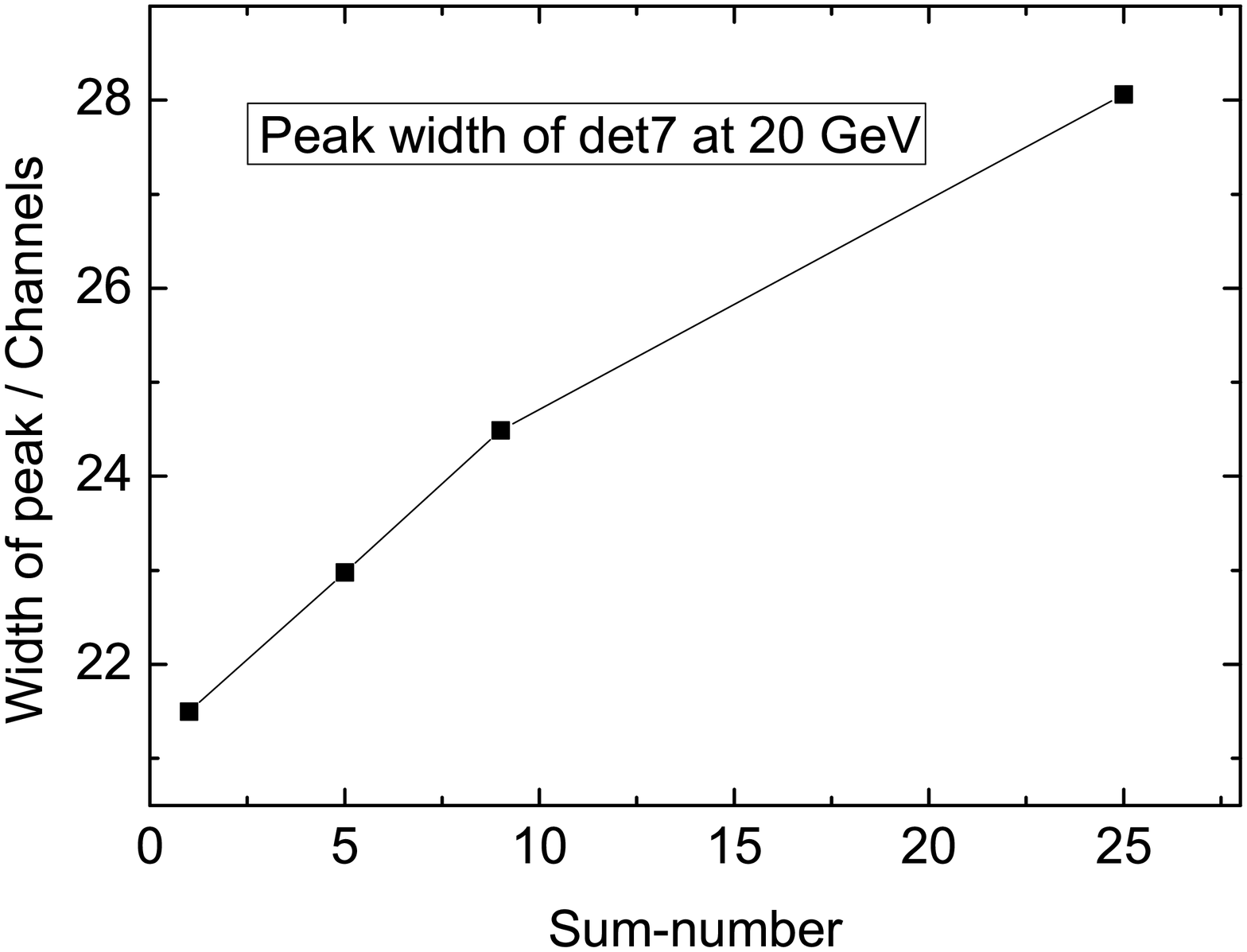}\\
\includegraphics[width = 0.4\textwidth]{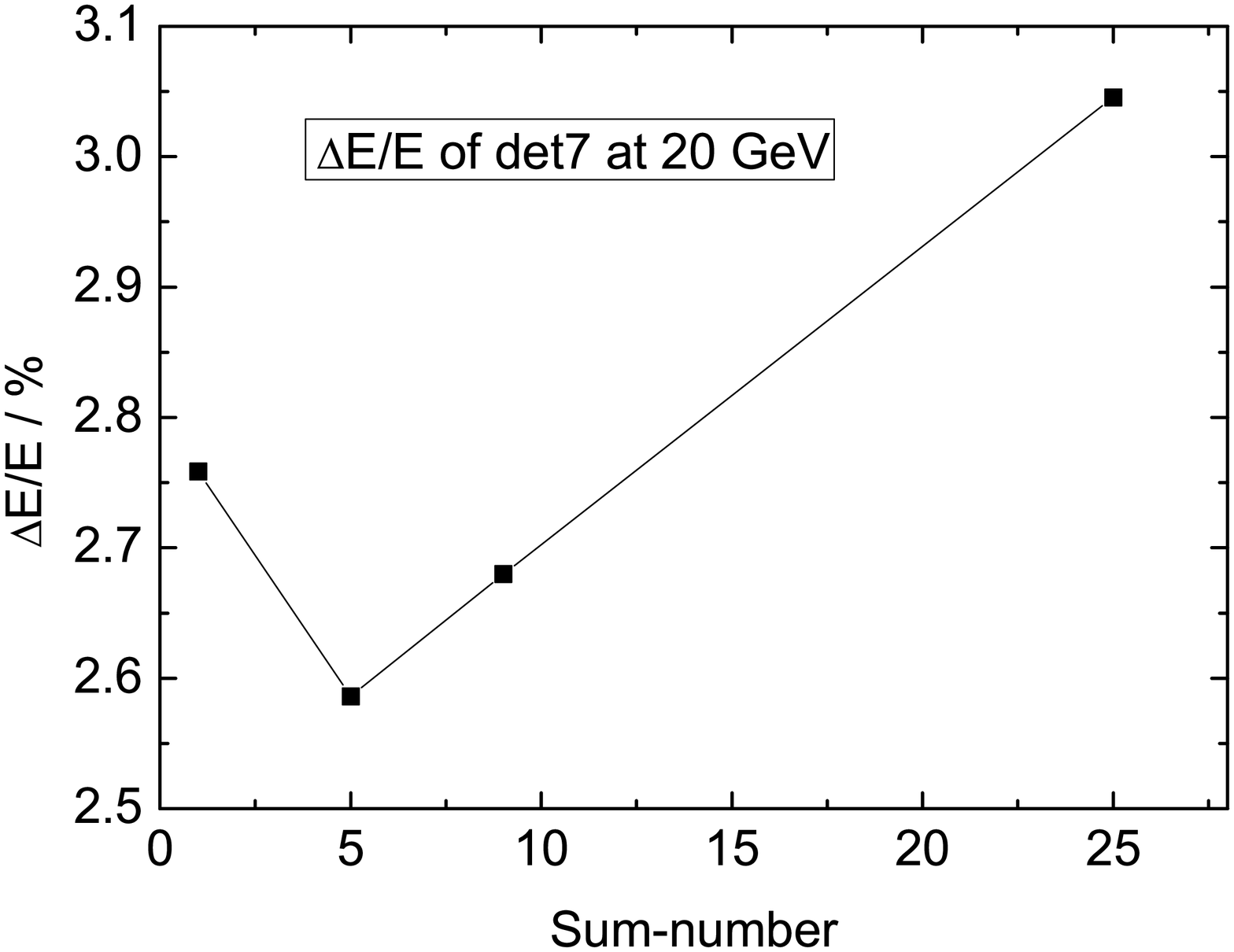}\\
\includegraphics[width = 0.4\textwidth]{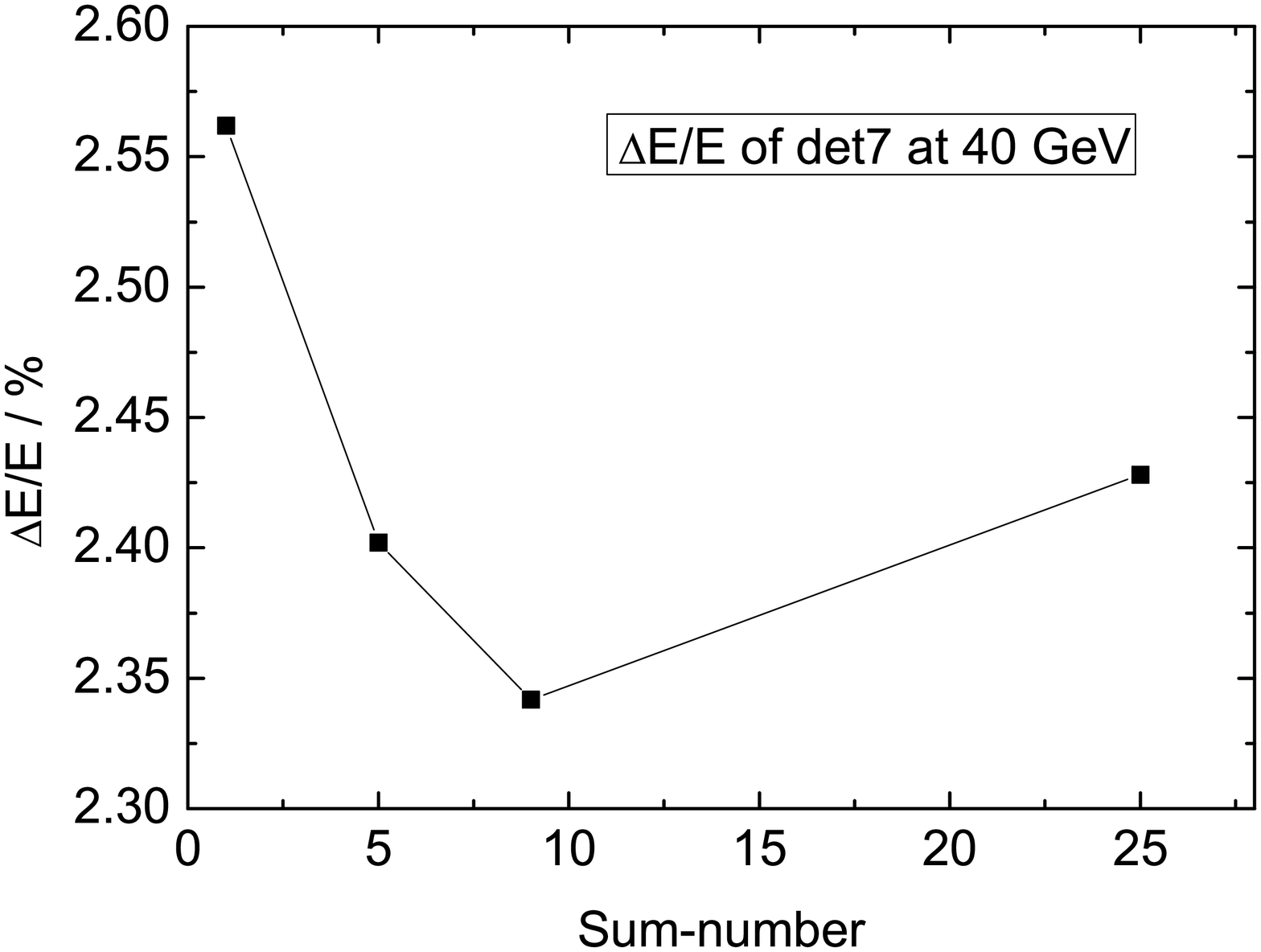}\\
\caption{\label{fig2}   The corrected energy peak position, peak width, and energy resolution of det7 changing with sum-number and beam energy.}
\end{center}
\end{figure}

Experimental data analysis shows that the corrected energy peak position increases and tends to a saturation value with the increase of the sum-number, and the width of the peak changes as well. Thus, the energy resolution is varying with the sum-number too. If the beam energy is higher, the sum-number should be bigger in order to get a better energy resolution. As it is shown in Figure \ref{fig2}.

Normally, the correction method with sum(3x3) or sum(1+4) is the best choice in the experimental energy range. 

\begin{figure}
\begin{center}
\includegraphics[width = 0.4\textwidth]{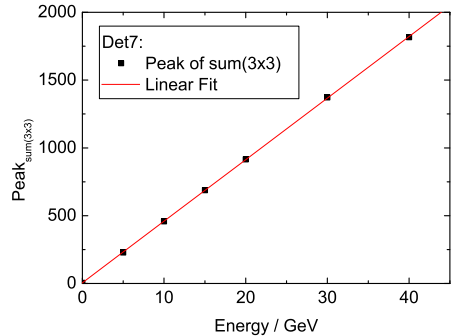}\\
\caption{\label{fig4} (Color online)  The linearity of the beam energy versus the peak position of det7.}
\end{center}
\end{figure}

The achieved linearity curve, i.e. the nominal beam energy versus the peak position, in ADC channels, is shown in Figure \ref{fig4}.
As can be seen from this figure, the response of the detector matrix is linear in a wide energy range 5-40 GeV. The adj. R-square of linear fit is 0.99996.
The result of the excellent linearity of the response of detector shows that the distortion of the signal caused by the punch-through effect is very small in the energy range. 
So, the punch-through effect can be ignored.

The energy resolution of an electromagnetic calorimeter can usually be parametrized as:
\begin{equation}\label{eq2}
  \frac{\sigma_E}{E}=\sqrt{\frac{a^2}{E^2}+\frac{b^2}{E}+c^2}
\end{equation}

Where the energy E is in GeV, $a$ represents the electronic noise, $b$ represents the stochastic term and $c$ represents the constant term. 
The noise term $a$ includes contributions from preamplifier noise, digitization noise and, in principle, pile-up noise, the latter was negligible in our tests.
The stochastic term $b$ takes into account the fluctuations in the electromagnetic showers and the variations due to photon statistics. 
The constant term $c$ takes into account shower leakage at the back end of the crystals, intercalibration errors, non-uniformity in the light collection and geometrical effects. 

\begin{figure}
\begin{center}
\includegraphics[width = 0.4\textwidth]{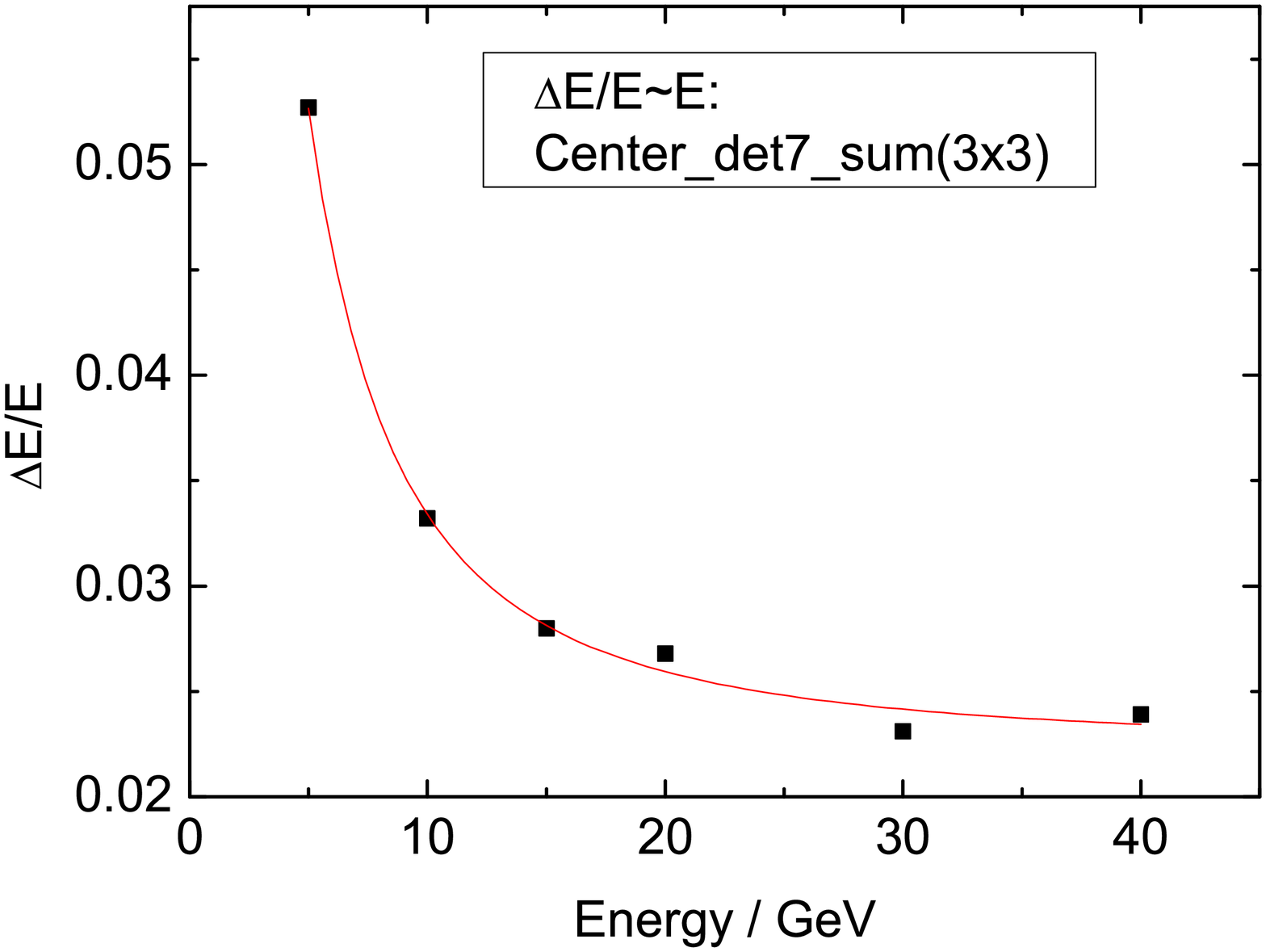}\\
\caption{\label{fig5} (Color online)  Energy resolution versus energy of the central det7 3x3 matrix, E = 20 GeV.}
\end{center}
\end{figure}

We have tested several 3x3 small matrixs with det7, det13, det17 and det19 (with Norwegian PINs) as the center respectively.
Table \ref{tab1} shows the energy resolution comparison of each central unit at E = 20 GeV.
The detector energy resolution with Chines PINs is equal to or a litter better than that with Norwegian PINs.

Figure \ref{fig5} shows the energy resolution curve versus energy of the central det7 3x3 matrix. 
In general, the energy resolution decreases while the energy increase.
The result of the fitting parameters with Formula (\ref{eq2})  are as follows: a = 227 MeV, b = 0.03, c = 0.02.

\begin{table}
\begin{center}
    \caption{\label{tab1} Comparison of energy resolution of central detectors, E = 20 GeV.}
    \begin{ruledtabular}
        \begin{tabular*}{120mm}{c@{\extracolsep{\fill}}cccc}
            Central detector & Det7 & Det13 & Det17 & Det19 \\
            \hline
            Energy resolution / \% & 2.68  & 2.59 & 3.03 & 3.25\\
      \end{tabular*}
     \end{ruledtabular} 
\end{center}
\end{table}

\section{Summary}

The beam test was performed of Chinese PIN diodes, assembled with PWO and preamplifier into a complete PHOS detector unit, at CERN SPS.
The energy resolution was measured with the electron beam energy ranging from 5 to 40 GeV.

In order to correct the energy loss of the electromagnetic shower into the surrounding detectors, to obtain the true energy peak and better energy resolution, the corrected method of summation was used.
The data analysis shows that the correction of sum(3x3) or sum(1+4) is the best choice in the experimental energy range.

The excellent linearity of the nominal beam energy versus the peak position of the detector shows that the distortion of the signal caused by the punch-through effect is negligible. 

An energy resolution comparison of each central unit at E = 20 GeV is given.
It shows that the detector energy resolution with Chines PINs is equal to or a litter better than that with Norwegian PINs.


\begin{acknowledgments}
This work is supported by National Natural Science Foundation of China (00121140488).
The authors would like to thank Prof. Mikhail Ippolitov, Prof. Arne Klovning, and the ALICE/PHOS-Collaboration for their kind help during the experiment at CERN. 
We thank the SPS staff for providing the experimental beam. 
We also thank Prof. Lu Zhang and his research group in Peking university for the collaboration during the R\&D of the PIN diododes.

\end{acknowledgments}

\end{document}